\documentclass[12pt, a4paper]{article}
\usepackage{cmap}
\usepackage[T2a, T1]{fontenc}
\usepackage[cp1251]{inputenc}
\usepackage[russian]{babel}
\usepackage[dvips]{graphicx}
\begin{document}
\title{Количественные методики восстановления истинных топографических свойств объектов по измеренным АСМ-изображениям:\protect\\
Часть 2. Эффект уширения АСМ-профиля}
\author{М.\,О.\,Галлямов, И.\,В.\,Яминский}
\date{\emph{Физический факультет Московского государственного университета им.\,М.\,В.\,Ломоносова, Москва}}
\maketitle
\begin{abstract}
Разработана методика количественного описания эффекта \emph{уширения} в АСМ, позволяющая восстановить реальные геометрические параметры объекта по двум измеренным параметрам АСМ-профиля (высота и ширина на полувысоте). Применение методики позволило получить информацию о количественном молекулярном составе комплекса ДНК-ПАВ.
\end{abstract}

Эффект уширения проявляется в том, что микрообъекты, визуализованные АСМ, имеют завышенные латеральные размеры. Например, при АСМ исследованиях молекул нуклеиновых кислот\,\cite{1} этот эффект облегчает идентификацию молекул: ``уширенные'' молекулы\footnote{Ширина профиля молекулы ДНК завышается в 5--10 раз } легче обнаружить на кадре значительной площади, что облегчает набор статистики. В силу этого, при исследовании нуклеиновых кислот, эффект уширения позволяет обходится без дополнительного контрастирования макромолекул (уранилацетатом и пр.).

Эффект уширения связан с тем, что зондирующее острие микроскопа имеет конечный радиус кривизны кончика. Эту аппаратную погрешность трудно преодолеть, поскольку уменьшение радиуса кривизны кончика зонда (использование более острых зондов) приводит к увеличению давления в области контакта (при том же значении величины контактных сил). Большее давление приводит к большим контактным деформациям зонда и образца, что сказывается на увеличении как латеральных размеров области контакта (ограничение достижимого пространственного разрешения), так и радиусов кривизны контактирующих поверхностей.

Уменьшить контактные деформации можно при наблюдении поверхности образцов в жидкостях, поскольку в этом случае можно поддерживать контактные силы на существенно более низком уровне\,\cite{2}. Однако, при исследованиях в жидкостях, могут возникать новые проблемы, среди которых мы упомянем проблему фиксации образца на твердой подложке.

Для восстановления реальной геометрической формы объекта по его АСМ-изображению необходим дополнительный математический анализ с использованием определенных модельных представлений о геометрии зонда (конус со сферическим кончиком, параболоид и пр.) и априорных представлений о форме объекта исследования.

В работе\,\cite{3} предложена универсальная компьютерная методика деконволюции АСМ-изображений, включающая два этапа: определение геометрии используемого острия с помощью тест-объектов и свертку инвертированной геометрии острия с измеренным АСМ-профилем; эта процедура позволяет во многих случаях восстановить исходный профиль объекта с высокой точностью. Мы проводили тестирование данной методики, решая задачу восстановления геометрии объектов, адсорбированных на поверхность плоской подложки. Анализ показал, что восстановленные по данной методике латеральные размеры объекта (ширина на полувысоте\footnote{этот параметр используется при определении объема исследуемого объекта}) существенно \emph{завышены} при условии, что радиус кривизны объекта \emph{меньше} радиуса кривизны кончика зонда (ошибка тем выше, чем больше разница соответствующих радиусов). Это обстоятельство осложняет применимость рассматриваемой методики при исследованиях биообъектов (макромолекул, их комплексов и пр.) в силу малости размеров последних в сравнении с радиусом кривизны кончика зонда АСМ.

В работе\,\cite{4} предложена методика восстановления объема исследуемых частиц (по АСМ-профилю), не включающая стадию предварительного тестирования зонда: как геометрия зонда, так и геометрия объектов исследования может быть восстановлена путем анализа одного и того же АСМ-изображения. Однако данная методика включает априорное предположение о сферической форме исследуемых объектов. Но, в силу имеющихся представлений о существенной роли \emph{контактных деформаций} в исследованиях АСМ представляется, что данное предположение вряд ли оправданно при решении задачи восстановления геометрии \emph{биообъектов}, характеризующихся, как известно, невысокими значениями модуля упругости. Более общей является модель, позволяющая учесть эффект уширения при контакте иглы с \emph{деформированной} частицей, имеющей \emph{эллипсоидальное} сечение. Но, насколько нам известно, в литературе отсутствуют работы, посвященные применению данной модели для анализа экспериментальных АСМ-изображений, что, по-видимому, обусловлено алгебраическими сложностями, возникающими при нахождении аналитического решения этой задачи.

\subsubsection*{Постановка и решение задачи о восстановлении реальной ширины объектов по измеренному АСМ-профилю}
Мы применили для учета эффекта уширения геометрическую модель (рис.\,1), учитывающую взаимодействие объекта только с кончиком зонда (предполагается, что нет контакта со стенками пирамиды). Это оправдано в том случае, когда высота исследуемых структур над подложкой не превышает радиуса кривизны кончика иглы.

\begin{figure}
\begin{center}
\includegraphics*[width = 0.7 \textwidth]{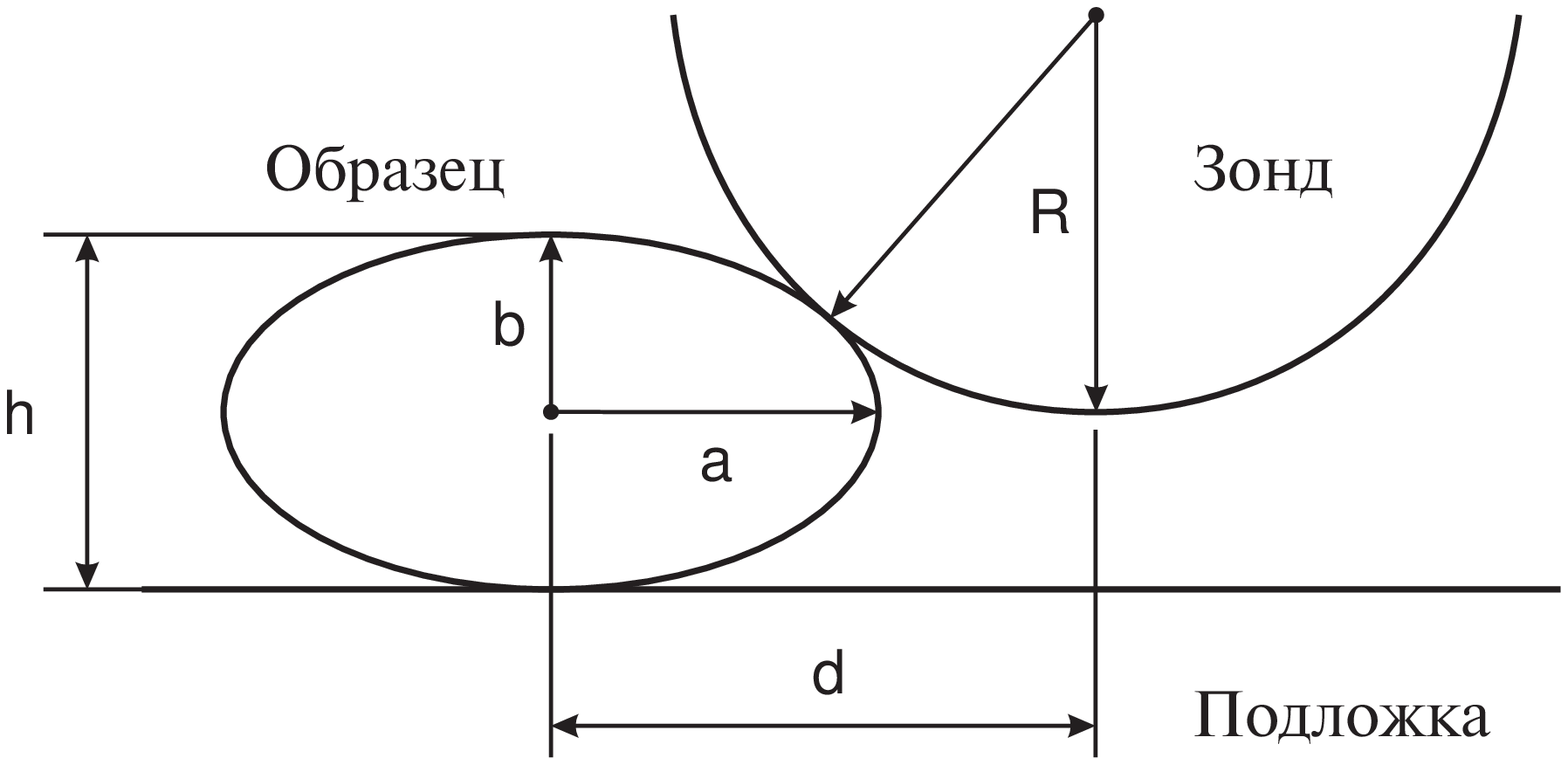}
\caption {Геометрия контакта зонда и образца (к объяснению эффекта уширения).
}
\end{center}
\end{figure}

Кончик зонда аппроксимировали либо полусферой радиуса $R$, либо параболоидом вращения (сечение иглы~--- парабола: $y=kx^2$, где $k$~--- коэффициент аппроксимации). При этом было показано, что результаты применения двух методик, фактически, тождественны. Количественное расхождение в ответах не превышает 3--9\%. Стоит отметить, что аппроксимация иглы с помощью полусферы более наглядна и более широко используется в литературе. Мы ниже приводим алгоритмы использования обоих подходов.

Исследуемая частица описывалась моделью сплюснутого эллипсоида. Т.е. мы исходим из априорных представлений о контактной деформации образца под действием зонда. В сечении объекта исследования~--- эллипс с полуосями $a$ и $b$. Ставилась задача по заданным значениям $b$ ($b = h/2$), $R$ (или $k$ для параболической аппроксимации) и $d$ (где $d$~--- измеренная ширина профиля АСМ-изображения частицы на ее полувысоте) найти значение $a$.

Запишем систему уравнений для эллипса (сечение профиля исследуемой частицы) и окружности (сечение профиля иглы):
$$
\left\{
\begin{array}{l}
y_{ell}=b\sqrt{1-x^2/a^2}\\
y_{cir}=R-\sqrt{R^2-(x-d/2)^2}
\end{array}
\right.
$$
В модели параболической иглы второе уравнение системы примет вид: $y_{par}=k(x-d/2)^2$.

Для решения задачи ставились условия для точки контакта с координатами ($x_0$, $y_0$): равенство касательных к эллипсу и к окружности (параболы) и удовлетворение координат точки контакта уравнениям эллипса и окружности (параболы):
\begin{equation}
\label{1}
\left\{
\begin{array}{l}
b\sqrt{1-x_0^2/a^2}=R-\sqrt{R^2-(x_0-d/2)^2}\\
dy_{ell}/dx|_{x_0, y_0}=dy_{cir}/dx|_{x_0, y_0}
\end{array}
\right.
\end{equation}
Для модели параболической иглы в системе\,(\ref{1}) вместо уравнения окружности используется уравнение параболы.
Данная система аналитически сводилась к одному уравнению с одним неизвестным ($x_0$)\footnote{для $x_0$ уравнение проще}, для которого был создан алгоритм численного решения\footnote{далее, по найденному $x_0$ определялось $a$}. Ниже мы приводим эти уравнения для случая сферической иглы:
\begin{equation}
\label{2}
\begin{array}{c}
\left[b^2-\left(R-\sqrt{R^2-(x_0-d)^2}\right)^2\right]\sqrt{R^2-(x_0-d)^2}+\\
+x_0(x_0-d)\left( R-\sqrt{R^2-(x_0-d)^2}\right)=0
\end{array}
\end{equation}
и для нахождения $a$ по численно определенному $x_0$:
\begin{equation}
\label{3}
a=\frac{x_0}{\sqrt{1-\left(R-\sqrt{R^2-(x_0-d)^2}\right)^2/b^2}}
\end{equation}
Аналогично, для случая параболической иглы имеем уравнение относительно $x_0$ для построения численного решения:
\begin{equation}
\label{4}
k^2(x_0-d)^3(x_0+d)+b^2=0
\end{equation}
и уравнение для определения $a$:
\begin{equation}
\label{5}
a=\sqrt{\frac{x_0(x_0+d)}{2}}
\end{equation}
Аналитически было показано, что полученное уравнение\,(\ref{2}) для сферической иглы не имеет единственного решения (на соответствующем промежутке) только в том случае, когда выполняется система неравенств:
\begin{equation}
\label{6}
\left\{
\begin{array}{l}
R>d/2\\
b>R-\sqrt{R^2-d^2/4}
\end{array}
\right.
\end{equation}
Аналогично, уравнение для параболической иглы\,(\ref{4}) не имеет единственного решения (на интересующем нас интервале) в случае, когда выполняется неравенство:
\begin{equation}
\label{7}
b>kd^2
\end{equation}
Смысл данных ограничения очевиден: если АСМ-профиль достаточно ``острый'', то и игла, с помощью которой он был прописан, также должна быть ``острой'' настолько, насколько это определяется соотношениями\,(\ref{6}) или\,(\ref{7}).

Уравнения (\ref{2}--\ref{5}) использовались нами для построения численного решения. Казалось бы, поскольку все равно задача решалась численно, то для реализации численного решения можно было исходить непосредственно из системы\,(\ref{1}), сведя ее, например, к системе нелинейных уравнений. Однако оказывается, что специфика системы не позволяет в этом случае построить достаточно простого численного решения, поскольку сколь угодно малые отклонения от точного решения приводят к отрицательным значениям в подкоренных выражениях, входящих в решаемую систему. В силу этого предложенный метод является, несмотря на необходимость предварительных аналитических выкладок, достаточно простым, кроме того, он имеет важное преимущество: проверка выполнения условий\,(\ref{6}) и\,(\ref{7}) позволяет заранее определить случаи отсутствия решений.

Эта проверка позволяет извлечь дополнительную и весьма важную информацию о свойствах зондирующего острия: определить верхнюю границу для значений радиуса кривизны кончика $R$ (соответственно, нижнюю для коэффициента параболы $k$).

Определение точного значения радиуса $R$ (или $k$) для конкретного зонда требует его тестирования непосредственно перед использованием (с помощью тест-объектов, например вирусных частиц\,\cite{5}). Однако и в этом случае существует вероятность того, что в процессе сканирования форма иглы претерпит изменения в результате взаимодействия с объектом. В этой связи чрезвычайно полезным является получение информации о форме зонда непосредственно из АСМ-изображений объекта исследования.

Путем набора статистики параметров высоты и ширины профиля АСМ-изображений объектов исследования и последующего анализа выполнимости соотношений\,(\ref{6}, \ref{7}) для набранной статистики можно определить предельное значение $R$ (или $k$) выше (или ниже) которого рост числа случаев отсутствия решения. Это и будет верхнее граничное значения для оцениваемого радиуса кривизны иглы, см.\,рис.\,2. Нижняя граница для радиуса кривизны зонда определяется контактными деформациями.

\begin{figure}
\begin{center}
\includegraphics*[width = 0.7 \textwidth]{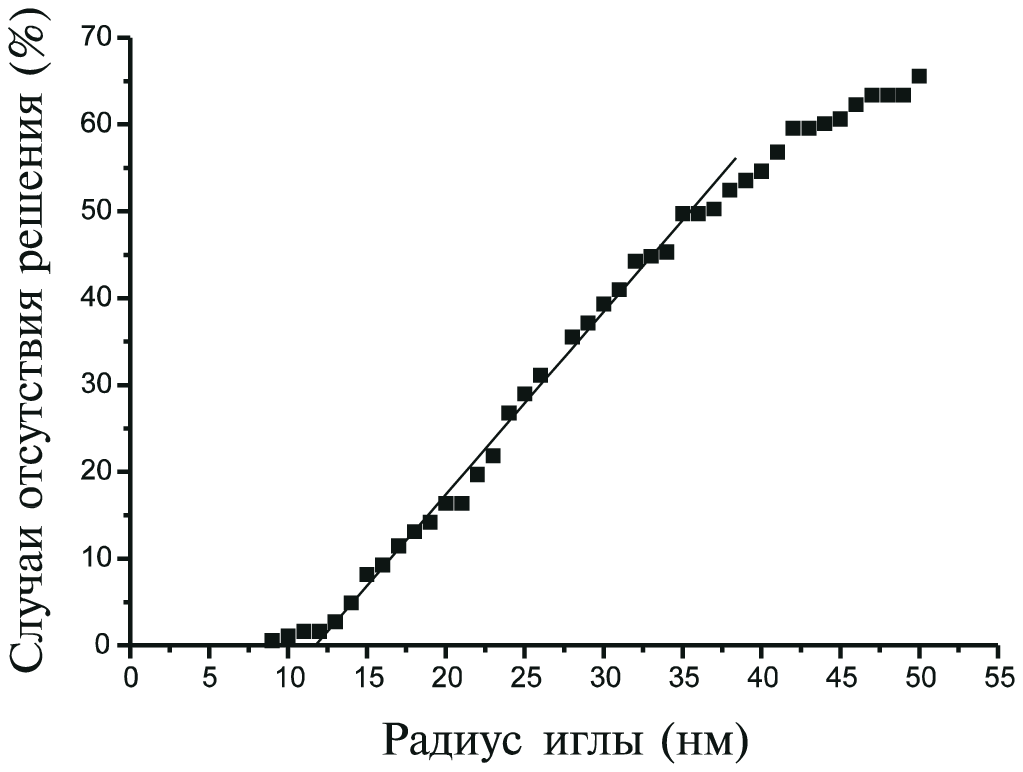}
\caption {Зависимость числа случаев отсутствия решения от радиуса аппроксимации иглы, проверка выполнимости условия\,(\ref{6}).
}
\end{center}
\end{figure}

\subsubsection*{К вопросу о применимости разработанной методики}
Следует подчеркнуть, что разработанная методика не учитывает возможного дополнительного вклада в уширение, связанного с частичным увлечением образца зондом при сканировании. Эффект увлечения обусловлен латеральными силами взаимодействия зонда и образца, которые характеризуются достаточно большой интенсивностью при проведении исследований в контактном режиме на воздухе, даже при минимизации нормальных сил. По нашим оценкам данный эффект, в основном, может проявляться при исследовании объектов, имеющих небольшую величину площади сечения (например, одиночных молекул ДНК), и приводить к 1{,}5--2{,}5-кратному завышению значения ширины объекта $d$ и, как следствие, восстанавливаемого значения $a$.
 
Степень увлечения образца зондом можно снизить при уменьшении интенсивности латерального силового воздействия зонда. Это достигается при применении режима прерывистого контакта. Максимальный эффект уменьшения латеральных сил достигается при измерениях в жидких средах.

\subsection*{Применение разработанного алгоритма для восстановлении морфологии комплексов ДНК-ПАВ}
Мы применили разработанную методику для восстановления геометрии комплексов ДНК с поверхностно-активными веществами (ПАВ), перешедших через границу раздела фаз вода/хлороформ, (подробнее результаты изложены в\,\cite{6}). Для каждой тороидальной частицы измерялись значения диаметра тора, ширины профиля на полувысоте и высоты над подложкой, средние значения этих параметров составили: $D\sim 100\,\mbox{нм}$, $d\sim 25\,\mbox{нм}$ и $h\sim 5\,\mbox{нм}$. Два последних значения использовались при восстановлении истинной ширины профиля частицы $2a$, по изложенной выше методике.

На рис.\,2 представлен график зависимости числа случаев отсутствия решения от радиуса аппроксимирующего зонда, полученный путем тестирования выполнения условий\,(\ref{6}) или\,(\ref{7}). На основании рисунка можно сделать вывод, что верхняя граница значения $R$, характеризующего зонд, используемый при визуализации комплексов, составляет 12\,нм (соответствует $k=5{,}5\times 10^{-2}\,\mbox{нм}^{-1}$)~--- выше этого значения наблюдается линейный рост числа случаев отсутствия решения.

\begin{figure}
\begin{center}
\includegraphics*[width = 1.0 \textwidth]{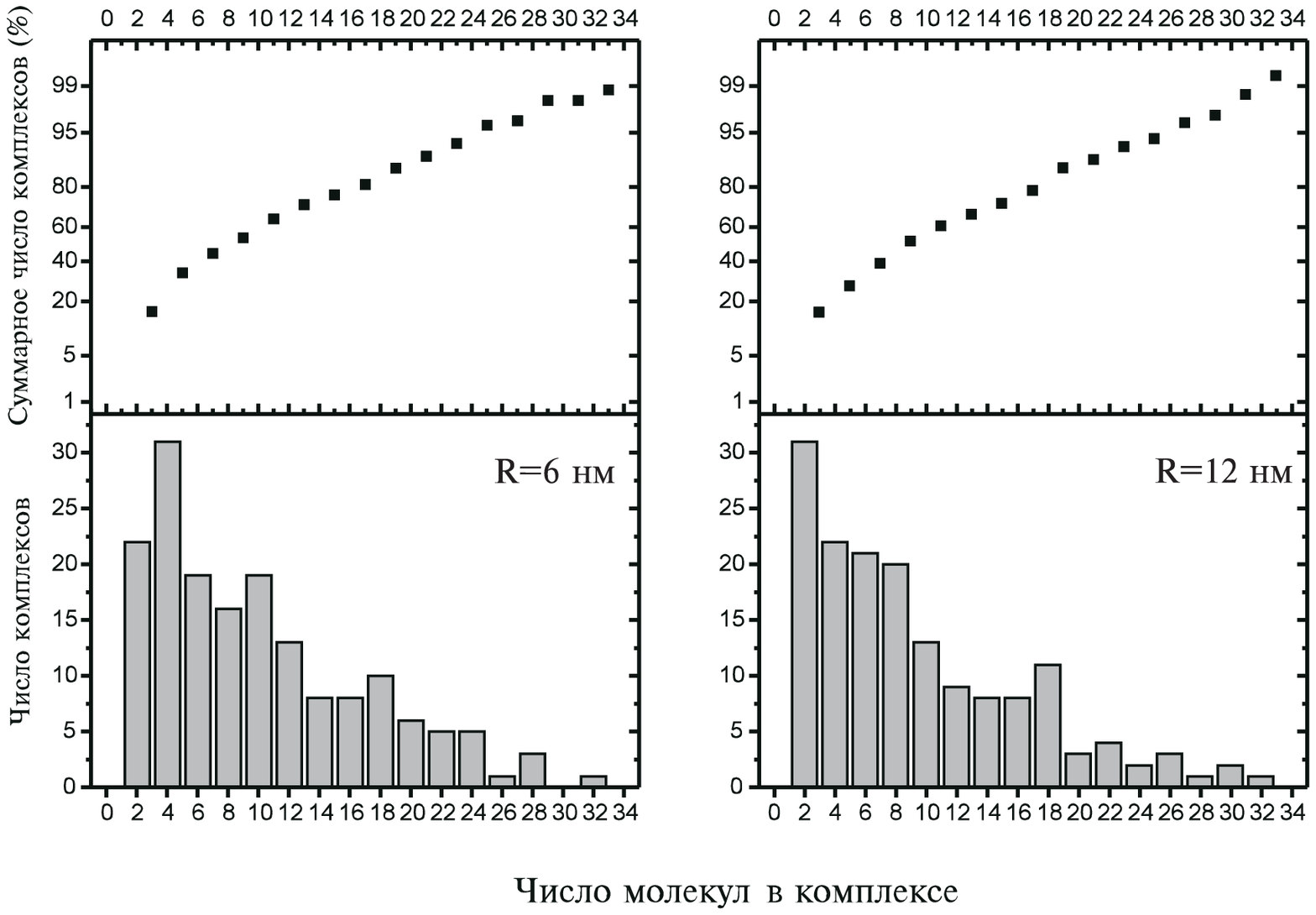}
\caption {Гистограммы распределения числа молекул ДНК, входящих в состав комплексов с ПАВ.
}
\end{center}
\end{figure}

Мы подсчитали количество молекул ДНК, приходящихся на каждую тороидальную структуру, используя два различных значения $R$~--- 6\,нм и 12\,нм (соответствует $k$~--- $13{,}2\times 10^{-2}\,\mbox{нм}^{-1}$ и $5{,}5\times 10^{-2}\,\mbox{нм}^{-1}$). Среднее значение параметра $a$ (найденное численно) для этих случаев составляет 11 и 10\,нм соответственно (относительное отклонение $\varepsilon =0{,}5$). Результаты применения параболической модели иглы дают, как правило, значения для $a$ на 3--9\% выше. В рассматриваемом случае соответствующие численные решения составляют 12 и 11\,нм. Таким образом, мы показали, что формой комплекса ДНК-ПАВ является сплюснутый тор. Восстановление геометрии комплекса позволяет провести количественный анализ его молекулярного состава, см. рис.\,3.

\emph{Благодарность}. Авторы выражают благодарность О.\,А.\,Пышкиной и А.\,С.\,Андреевой (Химический факультет МГУ) за помощь при приготовлении образцов комплексов ДНК-ПАВ. Работа была поддержана РФФИ, проект \No 97-03-32778a.

\section*{Quantitative methods for deconvolution of true topographical properties of object on the basis of measured AFM-images:\protect\\
Part 2. Broadening effect}
\subsection*{M.\,O.\,Gallyamov, I.\,V.\,Yaminsky}

Technique of the quantitative description of image broadening effect in AFM allowing to restore real geometrical parameters of object using two-parameters model (height and width at half-height) is developed. Application of the technique has allowed to receive the quantitative information about molecular structure of DNA-surfactant complexes.

\end{document}